\documentclass[aps,prl,preprint,superscriptaddress]{revtex4-2}
\usepackage{graphicx}
\usepackage{hyperref}
\usepackage{amsmath}
\begin{document}

\title{Unexpected detection rate dependence of the intrinsic detection efficiency in single-photon detectors based on avalanche diodes}

\author{Sebastian M. F. Raupach}
\email[]{sebastian.raupach@ptb.de}

\affiliation{Physikalisch-Technische Bundesanstalt (PTB), Bundesallee 100, 38116 Braunschweig, Germany}
\author{Ivo Pietro Degiovanni}
\affiliation{Istituto Nazionale di Ricerca Metrologica (INRIM), Strada delle Cacce 91, I-10135 Torino, Italy}
\affiliation{Istituto Nazionale di Fisica Nucleare (INFN), Sezione Torino, Via Giuria 1, I-10125 Torino, Italy}
\author{Hristina Georgieva}
\affiliation{Physikalisch-Technische Bundesanstalt (PTB), Bundesallee 100, 38116 Braunschweig, Germany}
\author{Alice Meda}
\affiliation{Istituto Nazionale di Ricerca Metrologica (INRIM), Strada delle Cacce 91, I-10135 Torino, Italy}
\author{Helmuth Hofer}
\affiliation{Physikalisch-Technische Bundesanstalt (PTB), Bundesallee 100, 38116 Braunschweig, Germany}
\author{Marco Gramegna}
\affiliation{Istituto Nazionale di Ricerca Metrologica (INRIM), Strada delle Cacce 91, I-10135 Torino, Italy}
\author{Marco Genovese}
\affiliation{Istituto Nazionale di Ricerca Metrologica (INRIM), Strada delle Cacce 91, I-10135 Torino, Italy}
\affiliation{Istituto Nazionale di Fisica Nucleare (INFN), Sezione Torino, Via Giuria 1, I-10125 Torino, Italy}
\author{Stefan K\"uck}
\author{Marco L\'opez}
\affiliation{Physikalisch-Technische Bundesanstalt (PTB), Bundesallee 100, 38116 Braunschweig, Germany}

\date{\today}

\begin{abstract}
Single-photon detectors are a pivotal component in photonic quantum technologies. A precise and comprehensive calibration of the intrinsic detection efficiency is of utmost importance to ensure the proper evaluation of the performance in view of the specific technological application of interest, such as the protection against security breaches in quantum cryptographic solutions. Here we report on a systematic study on and comprehensive analysis of the estimation of the intrinsic detection efficiency of two commercial single-photon detectors based on single-photon avalanche diodes (SPADs) for various mean photon numbers and at high laser pulse repetition rates using different techniques. We observed an unexpected and significant drop in intrinsic detection efficiency at detection rates of 10~\% and higher relative to the maximum detection rate. It is demonstrated that for data analysis a statistical model for the detection rate conveniently can be used if no time-stamped data are available. We conclude that the full characterization of single-photon detectors used in critical applications should include the sensitivity of their intrinsic detection efficiency to high event rates.
\end{abstract}


\maketitle

\section{Introduction}
Single-photon detectors based on single-photon avalanche diodes (SPADs) are a versatile tool for various applications ranging from quantum information and communication \cite{Mig04, Sas11, Pee09, Pirandola2020, Xu2020} to optical time domain reflectometry \cite{Med17}, biomedical research \cite{Kar17, Bru19}, time-of-flight laser ranging \cite{McC13, Ren11} and applications in the automotive industry \cite{Chan19}. These detectors are threshold detectors operating in Geiger mode. Consequently their responses inherently are highly non-linear with the mean number of photons \cite{Hadfield2009}.\\Applications typically aim at high count rates, e.g. to maximize the transmission rate in quantum communication \cite{Zha09, Dix08} or in satellite optical communication \cite{Liao2017, Val16}, or the sampling speed in laser ranging applications \cite{Du18}. \\The calibration of single-photon detectors' detection efficiency, defined as the probability that a photon impinging on the armed detector will produce a count, on the other hand typically is performed using strongly attenuated laser light at low count rates to avoid distortion effects in photon counting statistics due to unresolved multi-photon events and due to the holdoff time ('dead time') of the detector \cite{Had09, Tit19}. The calibration procedure consists in comparing the number of events registered by the single-photon detector per second to the mean incident power determined from a reference analogue detector. Typically a classical detection efficiency $\eta_{class}$ is modeled and determined, i.e. a response proportional to the impinging mean optical power \cite{lopez2015}. This is analogous to the substitution method employed in classical radiometry \cite{IEC61315,goebel1997}, and it does not account for the distortion effect on count statistics induced e.g. by the detectors' holdoff time and their inability to resolve the number of incident photons. The traditional approach usually assumes that these low-rate results are representative of the quantum behavior of the detector. This includes the tacit assumption that the intrinsic detection efficiency $\eta$ of the device, which is independent of the statistics of the incoming light, within the stated uncertainty is constant for all detection rates. Here, we scrutinize this assumption by measuring the intrinsic detection efficiency of commercial detectors based on InGaAs-SPADs, operated in free-running mode, using a pulsed laser over a broad range of mean photon numbers, focusing in particular on high event rates.\\A statistical model for the detection rate of such free-running detectors illuminated by pulsed sources was published recently \cite{georgieva2021}. The model was shown to describe well the change in 'click probability' and dark counts for mean photon numbers covering three orders of magnitude. It is applicable to setups where no information concerning the state of the single-photon detector with respect to its holdoff state is available, e.g. from a heralding event or an 'armed'-signal of the detector itself \cite{Brida2000, Brida2005, Castelletto2006, Min15}. Here we apply that model, and an improved version of it, to recover the quantum efficiency of single-photon detectors based on avalanche diodes, and compare the results to a software-based method that \textit{ex post} identifies and extracts only valid trigger events from time-stamped data . This latter approach should be an almost ideal one, as it is able to realize the definition of the intrinsic detection efficiency by eliminating the distortion effects due to the holdoff time. We consider only events within a few nanoseconds wide time window, which is small relative to the average dark count rate, thus minimizing the number of background counts mistaken as signal counts. In our measurements we find that, in the high rate regime, the quantum efficiency is not a constant parameter of the detector, but depends on the event rate. This suggests that future calibrations of the detection efficiency need to explicitly investigate this effect in order to avoid misconceptions which, among others, may induce security loopholes in quantum communication.  

\section{Setup and methods}

\subsection{Experimental setup}
\begin{figure}
\includegraphics[width=12.9cm]{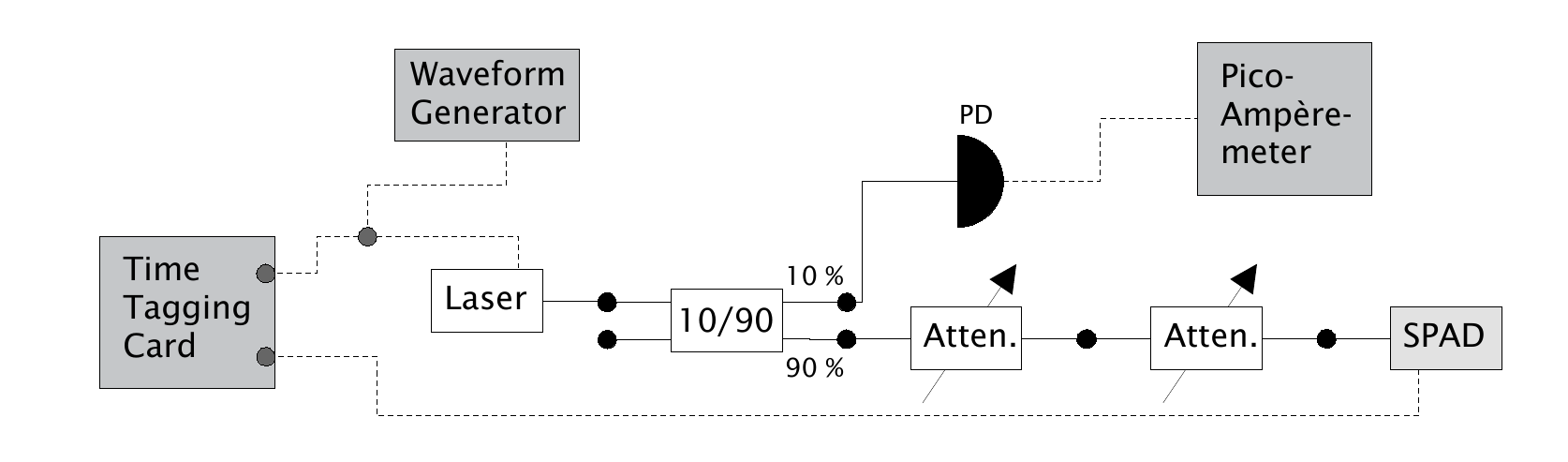}
\caption{Experimental Setup. A waveform generator referenced to PTB's 10 MHz standard frequency triggers a pulsed fibre-coupled laser (1548 nm, nominal pulse width $<$100 ps (FWHM)). After passing two variable attenuators, the laser pulse is detected by a single-photon detector based on an avalanche diode. Both the trigger pulses and the detector output signal are registered by a time tagging card.}
\label{fig:setup}
\end{figure}

To assess the statistical model-based approaches in retrieving the single-photon detector efficiency, we measured and analyzed the signal detection probability of two SPAD-based single-photon detectors having a nominal detection efficiency of $0.1$, and operated in free-running mode using a pulsed laser.\\The trigger signals to the laser were referenced to PTB's 10 MHz-standard frequency (see fig. \ref{fig:setup} for a sketch of the setup), and they were registered by a time tagging card. The laser is fibre-coupled and emits pulses at a nominal wavelength of 1548 nm with a nominal pulse width $<$ 100 ps (FWHM), where the setting of the mean power was kept constant for all measurements. A cooled, calibrated photodiode served as a power monitor. The diode's photocurrent $ i_\mathrm{mon}$ was measured and logged by a Femto-/Picoamp\`eremetre to calculate the mean photon number per pulse for each measurement run according to
\begin{equation}
n_\mathrm{ph} =\frac{ i_\mathrm{mon}\cdot q\cdot \alpha}{f\cdot s\cdot h\cdot \nu}.
\end{equation}
Here, $i_\mathrm{mon}$ is the photocurrent measured at a beam splitter's monitor output, $q$ is the ratio between the photocurrents measured at the monitor output and measured after the optical attenuator (at nominally 0 dB attenuation), $\alpha$ is its total linear attenuation (relative to nominally 0 dB), $f$ is the signal pulse repetition rate, $s$ is the calibrated sensitivity of the diode ($s= (1.0486\pm0.0063$) A/W ($k=1$)), $h$ is the Planck constant and $\nu$ is the optical frequency of the photons. The ratio of the photocurrents at the output of the second attenuator (both attenuators set to nominally 0 dB) and the monitor output of the fibre beamsplitter was  $q = 3.211$ (standard deviation $\sigma=0.008$). We found the combined relative standard uncertainty in mean photon number per pulse to be within the range $6.5\cdot 10^{-3}$ to $6.8\cdot 10^{-3}$ throughout the measurements, including only statistical variations for $i_\mathrm{mon}$ and $\alpha$ here.\\
The pulses passed an optical attenuation stage consisting of two fibre coupled variable attenuators in series. For our measurements, the total attenuation $\alpha$ covered a range from 56 dB to 72 dB.\\
We measured two commercial single-photon detectors based on InGaAs-SPADs (labeled as 'A', and 'B'), where the devices were nominally identical, i.e. both being of the same type and brand. For the measurements we employed user-set, nominal holdoff times of 10~$\mu$s and 20~$\mu$s. The background count levels, observed when the laser was not triggered, were around 820 counts/s and around 870 counts/s for devices A, and B, respectively, for a nominal holdoff time of 20~$\mu$s, and around 830 counts/s and around 910 counts/s for a nominal holdoff time of 10  ~$\mu$s.\\A time tagging card (Picoquant TimeHarp 260) attributed a time stamp with a nominal resolution of 250 ps to each event and streamed the events to hard disk for later analysis. For offline data analysis we used a software developed in-house.
\subsection{Data analysis}
For the retrieval of $\eta$, as a first approach we employed the statistical model presented in \cite{georgieva2021} (`original model') accounting for holdoff time and dark counts; here the signal detection rate $N_{click}$ is:
\begin{equation}
N_\mathrm{click} = f \frac{p_0}{1+m\cdot p_0}\exp\left(-N_\mathrm{dark,exp}D\right)+N_\mathrm{dark,exp}\left(1-\frac{p_0}{1+m\cdot p_0}\cdot f \cdot D\right),
\label{eq:model-overcycling}
\end{equation}
where $ N_\mathrm{dark,exp}$ is the darkcount rate measured in the absence of any signal pulses, $f$ is the pulse repetition frequency, 
\begin{equation}
p_0 = 1-\exp(-n_\mathrm{ph}\eta),
\label{eq:Poisson}
\end{equation} 
and $m$ is the integer part of the holdoff time $D$ divided by the pulse repetition interval. This takes into account 'overcycling', i.e. a repetition rate larger than the inverse of the holdoff time. The factor $\exp\left(-N_\mathrm{dark,exp}D\right)$ accounts for the probability of not having a dark count within the holdoff time prior to the detection event. According to eq. \ref{eq:model-overcycling}, $\eta$ can be estimated as:
\begin{equation}
\eta_\mathrm{orig.model} = -\frac{1}{n_\mathrm{ph}}\ln\left(1-\frac{a}{1-m\cdot a}\right),
\end{equation}
where
\begin{equation}
a = \frac{(N_\mathrm{click}-N_\mathrm{dark,exp})/f}{\exp\left(-N_\mathrm{dark,exp}D\right)-N_\mathrm{dark,exp}D},
\end{equation}
and $N_\mathrm{click}/f$ corresponds to the probability of observing a 'click' per laser pulse (click probability).\\In the following, we will observe that $\eta_\mathrm{orig.model}$ leads to a substantial overestimation in the regimes of high mean photon number per pulse and high repetition rate. For this reason, we amended the model in eq. \ref{eq:model-overcycling} by replacing the simple factor $\exp\left(-N_\mathrm{dark,exp}D\right)$ by an expression that also takes the effect of the signal detections in the estimation of the dark count rate into account (`amended model'). In this amended model the detection rate of eq. \ref{eq:model-overcycling} becomes:
\begin{equation}
\begin{split}
N_\mathrm{click} &= f \frac{p_0}{1+m\cdot p_0}\exp\left(-N_\mathrm{dark,exp}D\left(1-\frac{p_0}{1+m\cdot p_0}\cdot D\right)\right)\\
 &+N_\mathrm{dark,exp}\left(1-\frac{p_0}{1+m\cdot p_0}\cdot f \cdot D\right).
\end{split}
\label{eq:model-overcycling-adapted}
\end{equation}
It is not straightforward to estimate the quantum efficiency from eq. \ref{eq:model-overcycling-adapted} since this is a transcendental equation in $\eta$. Thus, we decided to solve it numerically.\\To validate the models for the estimation of quantum efficiency, where in both cases only lumped count rates are considered, we took advantage of each detection event being time-stamped individually. This allowed for establishing a software-based \emph{ex post}-validation of each trigger event to directly determine the 'true' click probability as a reference (`ground truth'). Using the detector's holdoff time as determined experimentally from several histograms for each detector, the software moved through the data stream in chronological order and flagged all trigger signals being 'shadowed' by the holdoff time of any previous detection event as being invalid. It then calculated the 'true' click probability as
\begin{equation}
p_{\mathrm{click, true}} = \frac{n_\mathrm{signal}}{n_\mathrm{trigger}-n_\mathrm{trigger,inv}}=1-e^{-n_\mathrm{ph}\eta},
\label{eq:groundtruth}
\end{equation}
where $n_\mathrm{signal}$ is the total number of detections within the signal detection window, $n_\mathrm{trigger}$ is the total number of registered trigger signals and $n_\mathrm{trigger,inv}$ is the number of triggers marked as being invalid. The \textit{ex post}-validation ensured that the non-linear detection effect induced by the holdoff time on the detection probability was eliminated. Furthermore, the effect of dark counts was largely suppressed as trigger pulses occuring during the holdoff time following a dark count are discarded as well.\\ In the following section we present experimental results for the retrieval of the single-photon detection efficiency $\eta$ according to the models given by eqs. \ref{eq:model-overcycling} and \ref{eq:model-overcycling-adapted}, as well as from the reference method based on \textit{ex post}-validation of trigger events.

\section{Results and Discussion}
\subsection{Measurements and retrieval of the quantum efficiency}
\begin{figure}
\includegraphics[width=17.2cm]{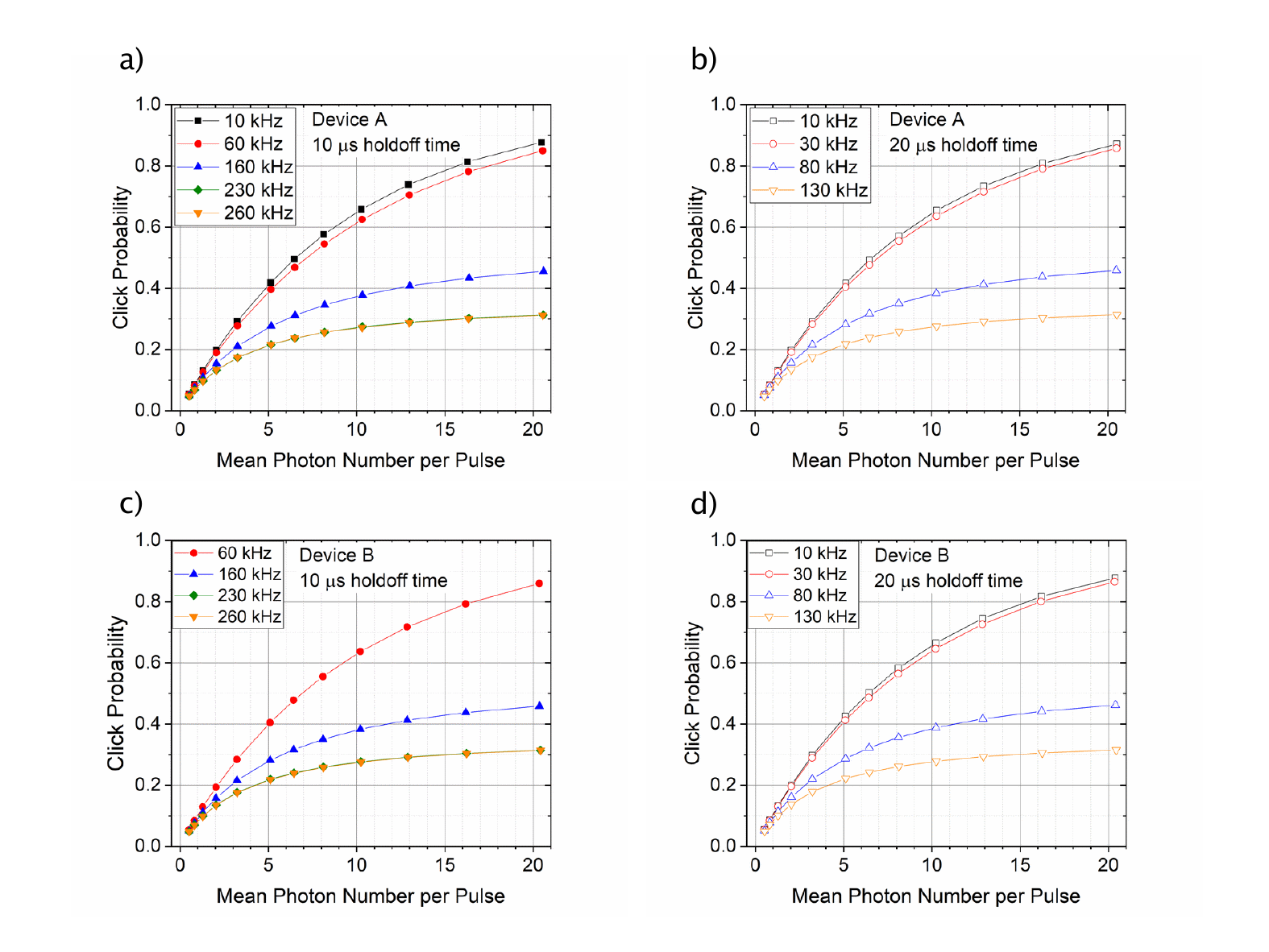}
\caption{Click probability; experimental data for both commercial SPAD-based single-photon detectors considered here. The graphs show the observed probability of a detection event within a 6 ns wide signal detection time window versus mean photon number per pulse, without corrections being applied.}
\label{fig:pclick}
\end{figure}
\begin{figure}
\includegraphics[width=17.2cm]{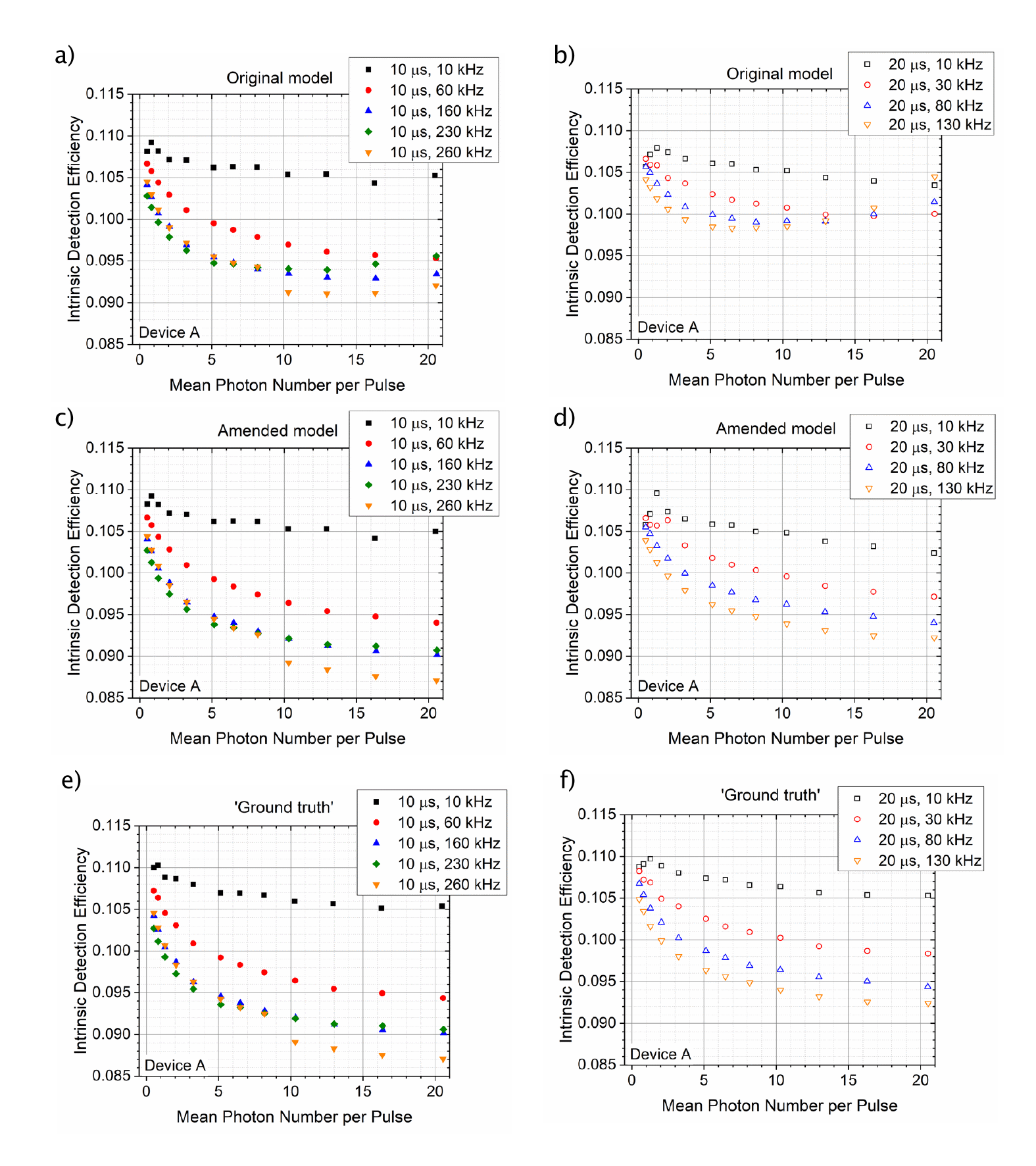}
\caption{Experimental data: Recovery of the intrinsic detection efficiency using the various methods for device A at nominal holdoff times of 10~$\mu$s (panels a, c, e) and 20~$\mu$s (panels b, d, f). }
\label{fig:eta-nphoton-A}
\end{figure}
\begin{figure}
\includegraphics[width=17.2cm]{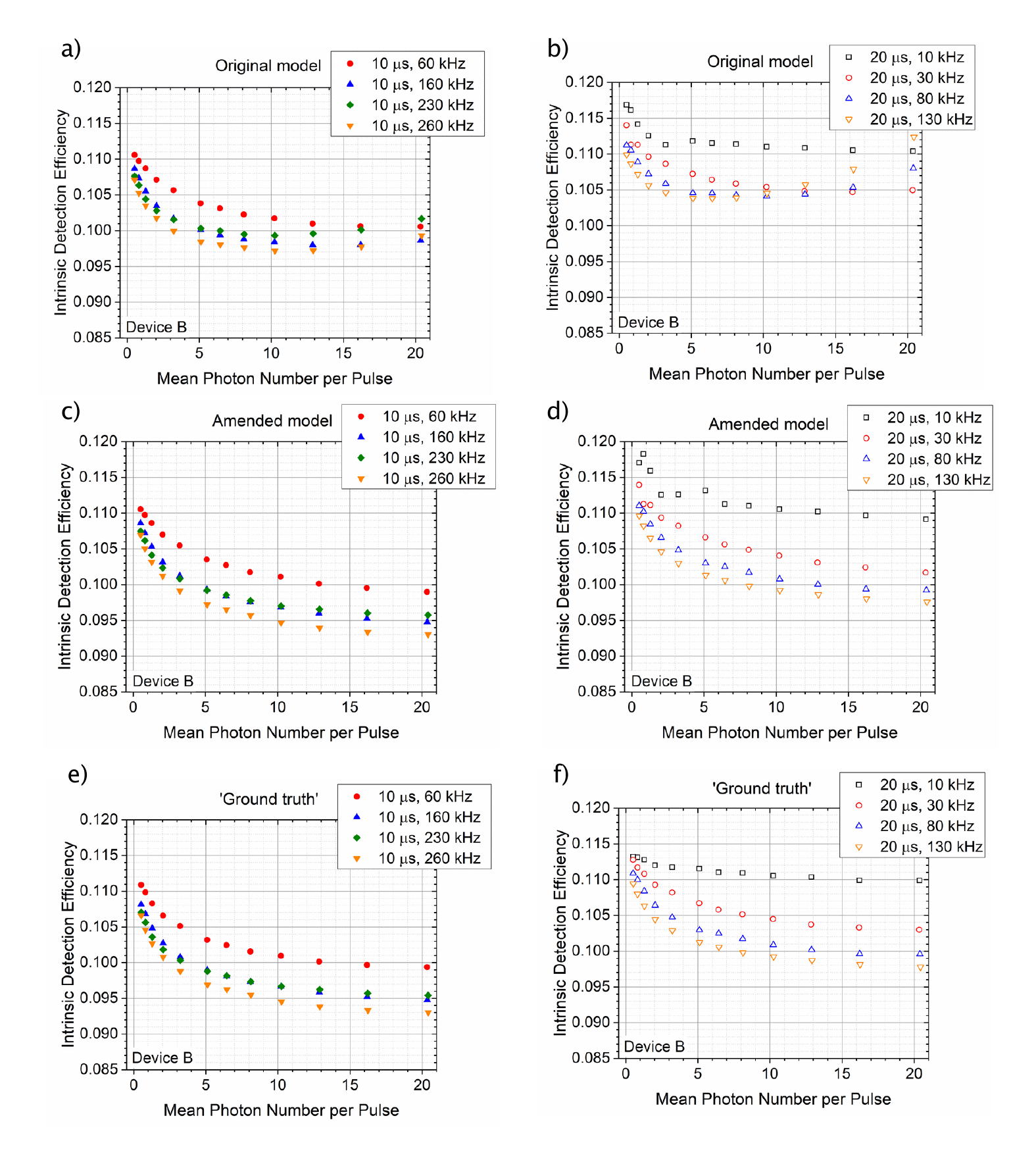}
\caption{Experimental data: Recovery of the intrinsic detection efficiency using the various methods for device B, at nominal holdoff times of 10~$\mu$s (panels a, c, e) and 20~$\mu$s (panels b, d, f).}
\label{fig:eta-nphoton-B}
\end{figure}

Figure \ref{fig:pclick} shows the observed mean click probabilities of the devices, i.e. the mean probability of registering an event within the signal detection window after registering a trigger, over a wide range of mean photon numbers and pulse repetition rates. Here, repetition rates below 100~kHz (50~kHz) correspond to a repetition interval larger than the dead time of around 10~$\mu$s (20~$\mu$s). The applied pulse repetition rates covered a range of the ratio $m$ of holdoff time ('dead time') to pulse interval of $m=0,1,2$ (see eqs. \ref{eq:model-overcycling}, \ref{eq:model-overcycling-adapted}, and \cite{georgieva2021}). The click probability, as expected, dropped significantly whenever the repetition rate became larger than an integer multiple of the inverse holdoff time. From the model, for a given $m$ the curves are expected to overlap. However, at e.g. 20~$\mu$s holdoff time and for repetition rates of 10~kHz and 30~kHz, both corresponding to $m=0$, a shift to lower click probability with higher repetition rate was visible (panels b) and d); similar in panel a), where for device A also for 10~$\mu$s holdoff time a measurement at 10~kHz repetition rate was done). This cannot be explained by the `overcycling' effect but hints at the presence of some other effect.\\Figures \ref{fig:eta-nphoton-A} and  \ref{fig:eta-nphoton-B} display the intrinsic detection efficiency (single-photon detection efficiency) $\eta$ retrieved for device A and B, respectively. The reconstruction of $\eta$ based on the original model of eq. \ref{eq:model-overcycling} (panels a and b) exhibits a visible upward slope with increasing mean photon number and increasing repetition rate compared to the reconstructions based on the amended model (eq. \ref{eq:model-overcycling-adapted}, panels c and d) as well as compared to the reference values (`ground truth', panels e and f). This underlines that the effect of pulse rate and signal detection probability on the background count rate needs to be included into the statistical model as is done in eq. \ref{eq:model-overcycling-adapted}.\\
However, most notably we found that the values of $\eta$ spread out and in general show an overall decrease with increasing mean photon number and with increasing pulse repetition rate. In particular this was also true for the values of $\eta$ estimated from the \emph{ex post}-validated data. To assess if this is an artefact of the reconstruction, e.g. due to afterpulses not being accounted for in the models, a closer look at the devices' background counts is required.

\clearpage
\subsection{Assessing the  background count dynamics}
\begin{figure}
\includegraphics[width=17.2cm]{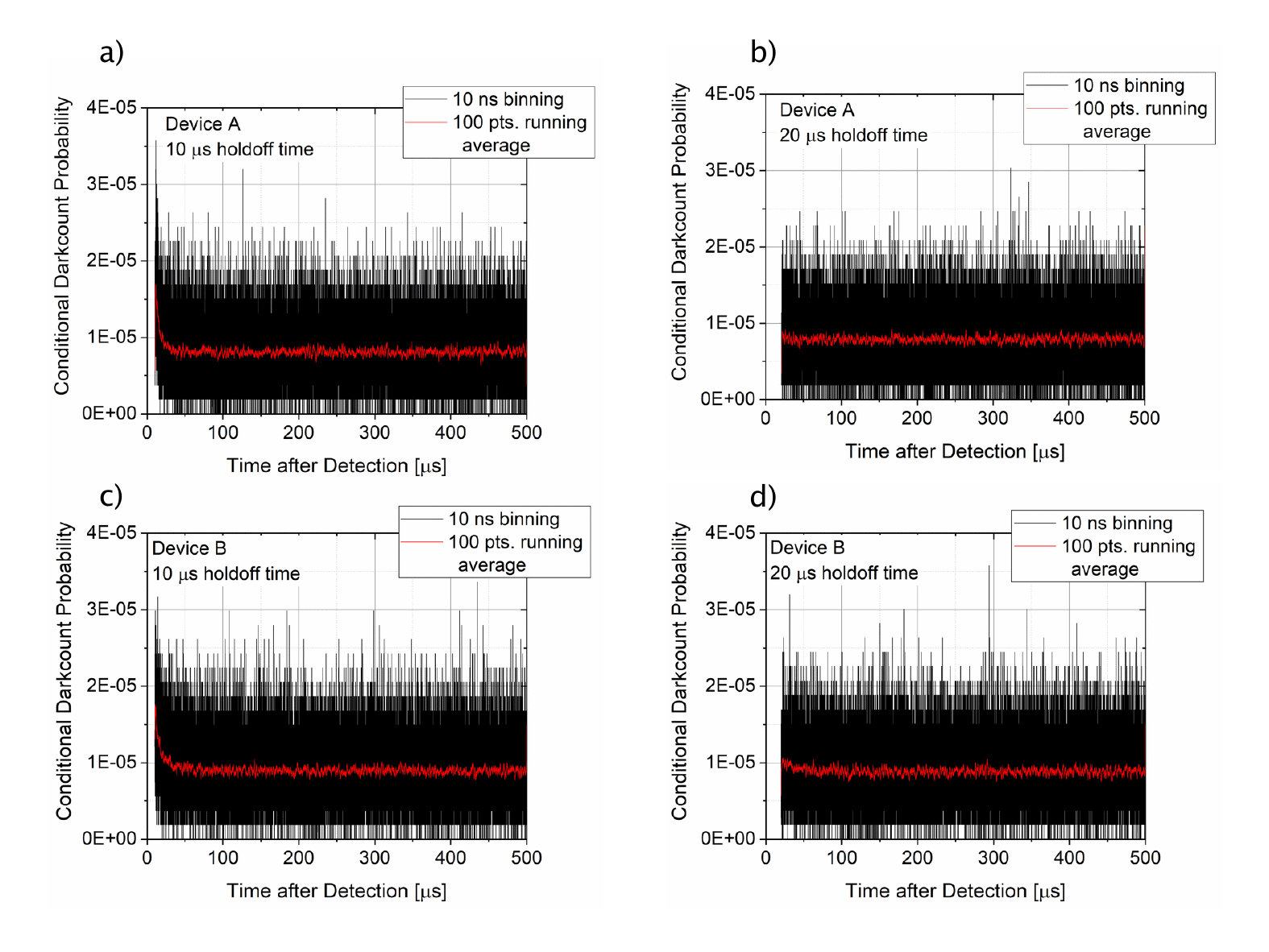}
\caption{Histograms of the dark count probability conditional on a successful signal detection, at a nominal holdoff time of 10~$\mu$s (panels a,c) and 20~$\mu$s (panels b,d) for the single-photon detectors considered here (10 ns bins). All data were taken at a pulse repetition rate of 2~kHz and a mean photon number of around 20 photons/pulse (measurement time 120 s each).}
\label{fig:histograms}
\end{figure}

\begin{figure}
\includegraphics[width=17.2cm]{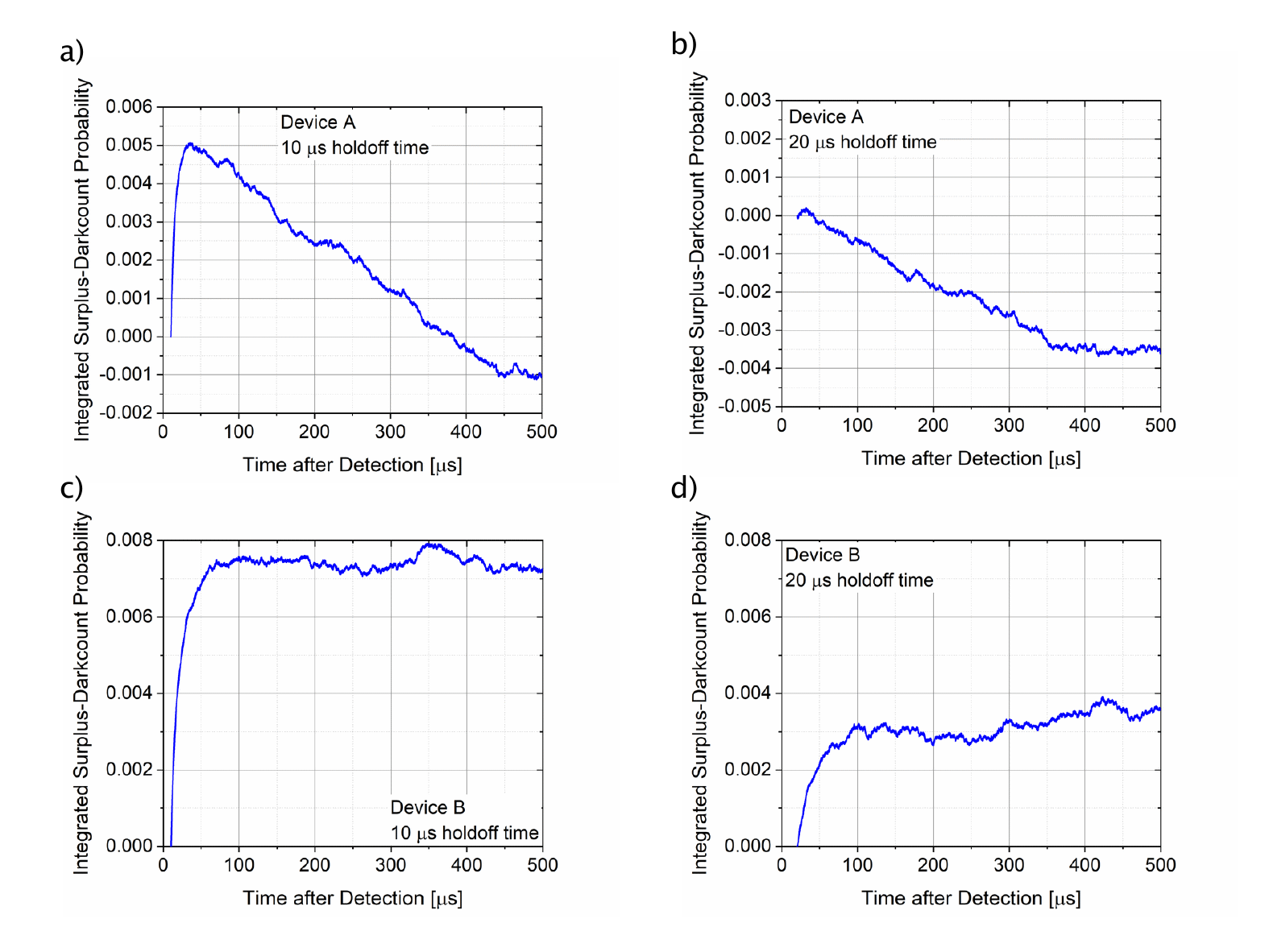}
\caption{Sum of the surplus-darkcount probability, e.g. due to afterpulsing, at a nominal holdoff time of 10~$\mu$s (panels a,c) and 20~$\mu$s (panels b,d), calculated over the time elapsed since a detection event within the signal detection time window (see figure \ref{fig:histograms}).}
\label{fig:surplusdarkcounts}
\end{figure}

As the models do not include dynamic changes of the background counts such as afterpulsing, one might conclude that the devices' dark count dynamics causes the observed behaviour. To characterize these dynamics, we performed a measurement at a low repetition rate of 2~kHz and at a high mean photon number of around 20 photons per pulse. Figure \ref{fig:histograms} shows histograms of the dark count probability for each of the devices for time bins with a width of $\Delta t = 10$ ns. We considered only events that occured within the repetition interval of 500~$\mu$s after each trigger, filtered for triggers where an event was detected within a given 6 ns wide signal detection window ('successful trigger'). We divided the total number of events in each bin by the number of signal detections to obtain the conditional dark count probability. 
As is visible in a qualitative manner already from the histograms of this (conditional) dark count probability, both device A and B exhibited a rather low afterpulsing in particular for nominal holdoff times of 20~$\mu$s. In general, when applying the original or the amended model to time-unresolved data, the conditions of operating the single-photon detector should be chosen such that afterpulsing is minimized as it is not included into the models. However, in most situations, and also in the measurements presented here, it cannot be avoided entirely. Consequently, our measurements did include the effect of very small afterpulsing.\\
For visualizing the dark count dynamics following a signal detection, we first calculated the mean value $\bar{p}_{\mathrm{dc}}$ of the dark count probabilities over the last 5000 bins (50~$\mu$s) of each histogram as a long-term reference value. We subtracted $\bar{p}_{\mathrm{dc}}$ from each bin and summed up the residual probability over the time bins to arrive at the integrated surplus-darkcount probability relative to the respective reference value:
\begin{equation}
P_{\mathrm{dc,surplus}}(t) = \sum_{i = 0}^{i=t/\Delta t} \left(p_{\mathrm{i,dc}}(\Delta t)-\bar{p}_{\mathrm{dc}}\right)
\end{equation}
 An elevated count level e.g. due to afterpulsing, will increase this quantity with integration time, while a count level lower than the respective reference value will lead to a decrease of the integrated surplus-darkcount probability. Counts fluctuating around the reference value will on average not contribute to the sum.\\The integrated surplus darkcount probability, fig. \ref{fig:surplusdarkcounts}, allowed for a more quantitative comparison as well as for detecting subtle differences in the dynamics between device A and B. The surplus dark count probability for device B leveled off at a click probability of less than 0.8~\% (less than 0.3~\%) after less than 100~$\mu$s for a nominal holdoff time of 10~$\mu$s (20~$\mu$s). Device A exhibited a maximum surplus-darkcount probability of 0.5~\% ($<0.05~\%$) at a nominal holdoff time of 10~$\mu$s (20~$\mu$s) followed by a steady decrease. This decrease indicates that the dark count probability on average was below the mean value of that in the last 50~$\mu$s of the 500~$\mu$s interval. This is consistent with observations made separately, where we saw a suppression of background counts for that device after a detection event, followed by a long-term 'recovery' of the background count level. The surplus-darkcount probability is a convenient means to visualize such subtle differences in background dynamics e.g. when comparing nominally identical devices.\\
\clearpage
\subsection{Retrieving $\eta$ from simulations}
\begin{figure}
\includegraphics[width=17.2cm]{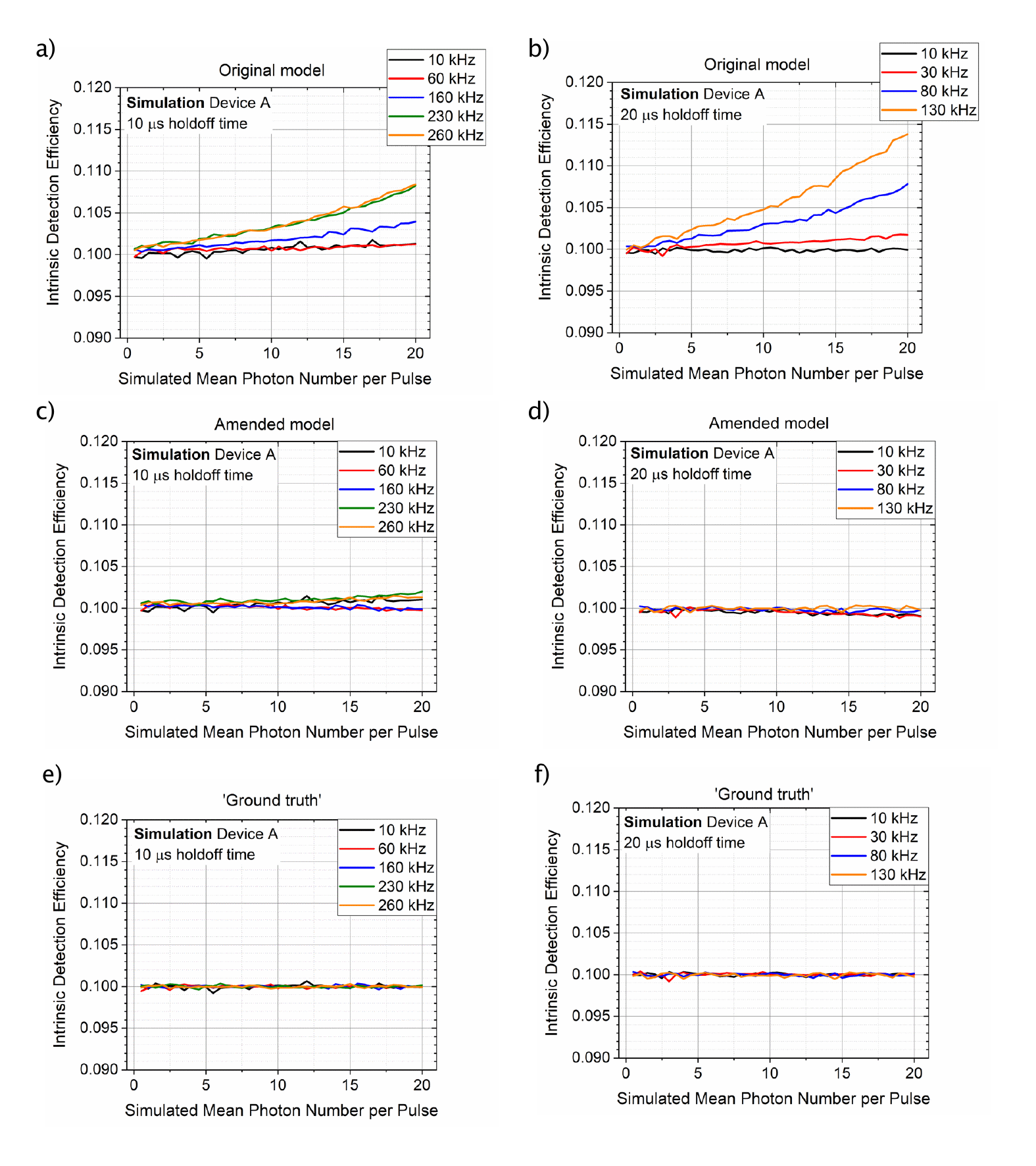}
\caption{Simulation, device A: Intrinsic detection efficiency recovered with the various reconstruction methods from a simulated stream of detection events, where the simulation of dark counts is based on the dark count probabilities shown in figure \ref{fig:histograms} (here: $\eta = 0.100$).}
\label{fig:simulationA}
\end{figure}

\begin{figure}
\includegraphics[width=17.2cm]{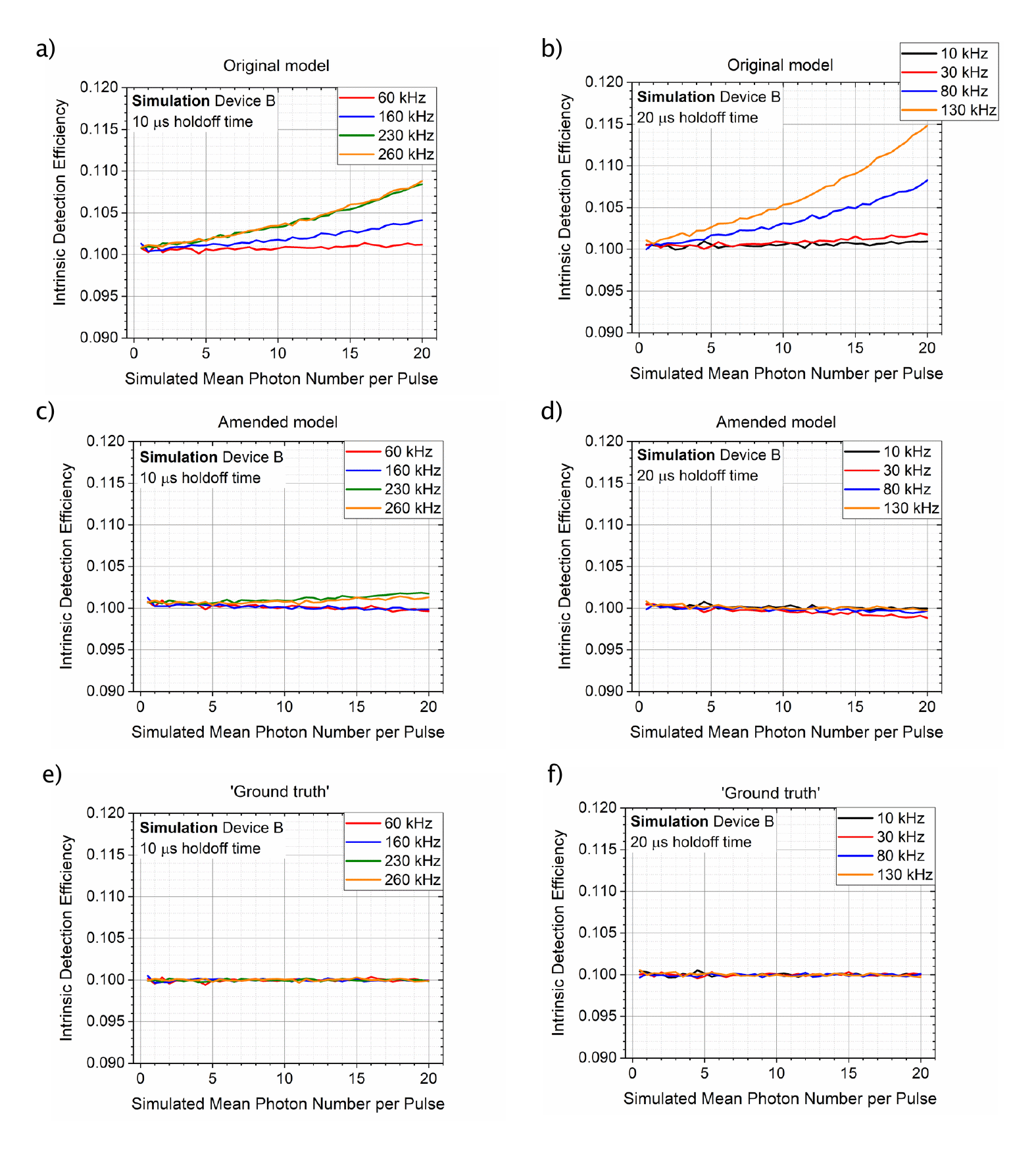}
\caption{Simulation, device B: Intrinsic detection efficiency recovered with the various reconstruction methods from a simulated stream of detection events, where the simulation of dark counts is based on the dark count probabilities shown in  figure \ref{fig:histograms} (here: $\eta = 0.100$). }
\label{fig:simulationB}
\end{figure}

To further assess the fidelity of retrieving the intrinsic detection efficiency using the models as well as the \emph{ex post}-validation, in particular in view of the observed background count dynamics, we additionally performed an analysis of simulated data streams. To obtain a realistic scenario, we made use of the dark count dynamics observed experimentally. In the simulation, the probability of a dark count occuring after a signal 'detection' was governed by the dark count probabilities shown in fig. \ref{fig:histograms}. Hence, weak afterpulsing and even dark count suppression was included into the simulations.\\We set the simulated detector's efficiency to exactly $\eta=0.1$, and iterated in time steps of 10 ns. For each step, we checked if a multiple of the simulated signal pulse repetition interval was reached. If not, we allowed for the possibility of a dark count according to the corresponding time bin of the experimental dark count probability (for larger time intervals than 500~$\mu$s, again the mean value of the last 5000 histogram bins was used as a constant probability value). If, on the other hand, the time step reached a multiple of the simulated pulse repetition interval, we first allowed for the possibility of a signal detection to occur with a probability of  $p_0 = 1-\exp(-n_\mathrm{ph}\eta)$ before allowing for the possibility of dark count ($n_\mathrm{ph}$: simulated mean photon number per pulse). For each mean photon number, we simulated the data streams over a total number of time steps corresponding to a total time of 100 s. \\The results of retrieving $\eta$ from the simulated data are shown in figs. \ref{fig:simulationA} and \ref{fig:simulationB}. We found that despite low afterpulsing probabilities, see fig. \ref{fig:simulationA}, panels a) and b), the reconstruction of the intrinsic detection efficiency based on original model from \cite{georgieva2021} indeed tended to overestimate $\eta$ for all but the lowest repetion rates and/or low mean photon numbers, where this overestimate increased with mean photon number and pulse repetition rate.  The amended model, on the other hand, which takes the backaction of signal detections on the dark count rate into account, largely suppressed the deviation exhibited by the original model. For very small mean photon numbers both models yielded largely similar results. When applying the \emph{ex post}-validation of the trigger pulses followed by the  evaluation of eq. \ref{eq:Poisson}, the correct value of $\eta$ was recovered faithfully for all repetition rates and mean photon numbers. We verified the insensitivity of the \emph{ex post}-validation to background count dynamics further by also performing simulations based on histograms obtained from a different device suffering from strong afterpulsing (including higher-order afterpulsing).  Also in these simulations, an \emph{ex post}-validation allowed for a successful retrieval of $\eta$, very similar to the simulations shown here. We therefore conclude that the results obtained via an \emph{ex post}-validation are not affected significantly by afterpulsing but yield a meaningful reference value ('ground truth') for $\eta$, when applying the models to experimental data.\\Therefore, we are led to conclude that, unlike the upward slope of $\eta$ for the original model, the experimentally observed spread and variation of $\eta$ is not an artefact of the reconstruction but a property of the devices under test. Contrary to the general assumption, their intrinsic detection efficiency $\eta$ at least in the range covered here cannot be considered as being a constant.
\\

\clearpage
\subsection{Time-interval dependence of the quantum efficiency}
\begin{figure}
\includegraphics[width=17.2cm]{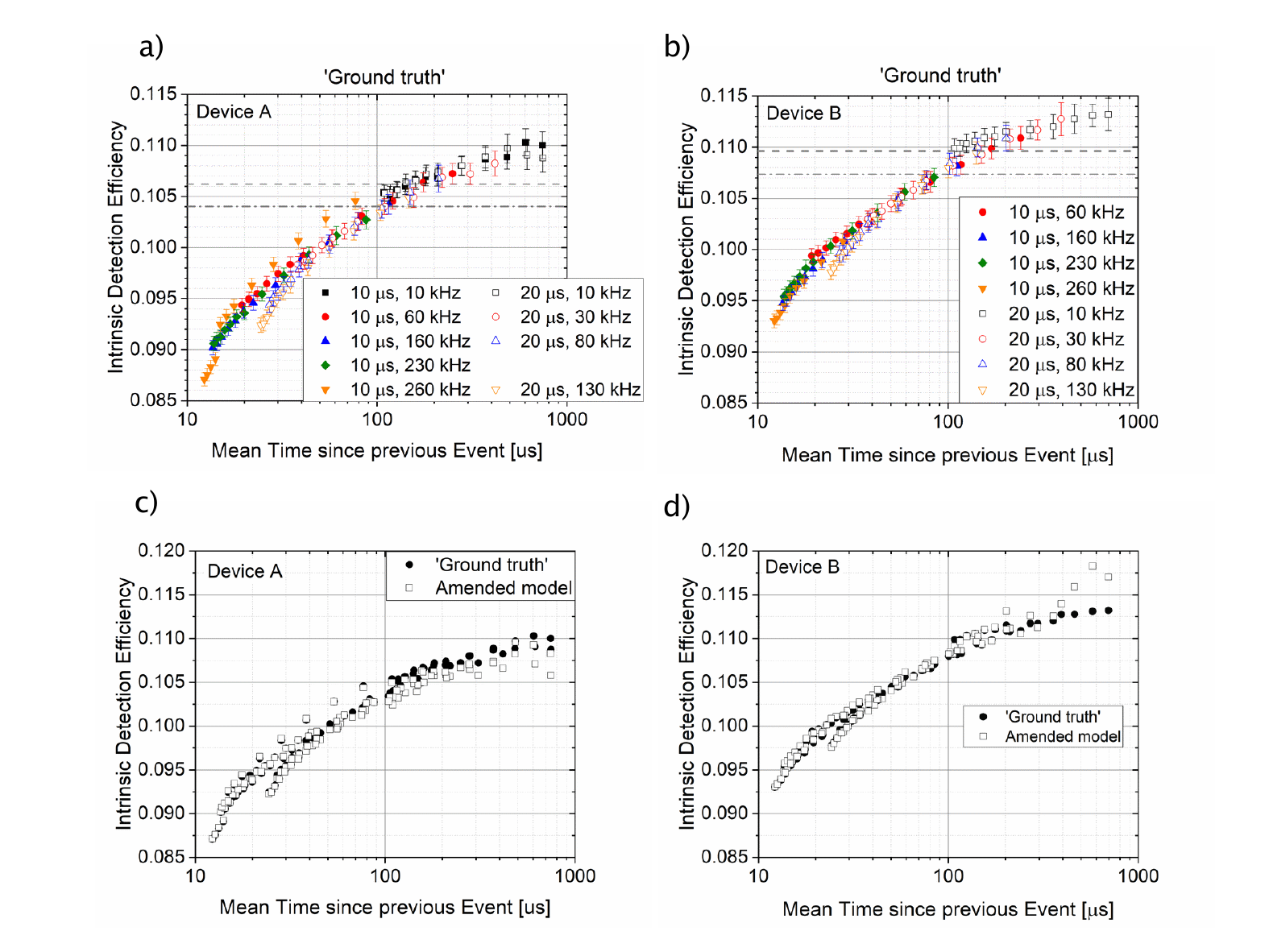}
\caption{Displaying the experimental values for the intrinsic detection efficiency (retrieved after \emph{ex post}-validation of the time-stamped data) versus the mean time to the previous event for each measurement on a semi-logarithmic scale. Panels a) and b) display $\eta$ over mean time to the previous event for the signal detections by device A and B, respectively. The dashed (dash-dotted) line indicates a relative decrease of approximately 3~\% (5~\%) relative to the value recovered for the intrinsic detection efficiency at the longest inter-detection time intervals. The relative standard deviation of the mean time since the  previous event is $<3.8\cdot 10^{-3}$ for all cases. Panels c) and d) displays $\eta$ recovered from the amended overcycling model in comparison to the 'ground truth' from the \textit{ex post}-validated data. }
\label{fig:eta-deltat}
\end{figure}

As the variation of $\eta$ with mean photon number and repetition rate lends itself to the idea of a possible time dependence, we tentatively plotted the estimated values of $\eta$ retrieved from the \textit{ex post}-validated data over the mean time interval between a signal detection event to the last preceding event (background count or signal), see fig. \ref{fig:eta-deltat}. In particular at high rates, where the background counts are negligible (see also \cite{georgieva2021}), this essentially corresponds to the inverse of the mean detection rate. The error bars indicate the combined standard uncertainty of $\eta$. The observed relative standard deviation of the mean time interval monotonically grows with decreasing mean photon number for both devices to values of up to $3.8\cdot10^{-3}$. \\
To assess the significance of the variation in the recovered intrinsic detection efficiency, we refer to a recent international pilot study regarding the calibration of a SPAD-based single-photon detector's efficiency \cite{lopez2020}, which stated expanded relative uncertainties between 2.7~\% and 5.3~\%. We include a dashed (dash-dotted) line indicating a decrease of $\eta$, relative to its long-interval value by around 3~\% (around 5~\%) .\\
For both detectors, we found that the reconstructed values of $\eta$ largely overlap and follow a common curve, where the intrinsic detection efficiency decreased with decreasing mean inter-detection time interval (increasing detection rate). Here, the decrease in observed quantum efficiency became significant for mean time intervals below 100~$\mu$s, roughly corresponding to detection rates around 10~kHz. For a nominal holdoff time of 10~$\mu$s, this corresponds to about 10~\% of the maximum detectable event rate. We note that at this rate the repetition intervals were much larger than the holdoff time, i.e. the observed effect was not due to pulses sampling the partial recovery of the device at the end of its holdoff time. Towards the respective holdoff time we found a steepening slope, leading to a kind of 'bifurcation' between the 10~$\mu$s and the 20~$\mu$s data below 30~$\mu$s. This may possibly be a hint to the mechanisms that lead to the observed behaviour, or may indeed be due to sampling the mentioned recovery period. When comparing the values of the quantum efficiency as reconstructed using the amended model to the reference values of $\eta$ retrieved after \textit{ex post}-validation of the data, we found that they substantially overlap. For both approaches we consistently observed a significant variation of the intrinsic detection efficiency, which would not have been noted in measurements at low event rates.

\section{Conclusion}
In this paper, using simulated and experimental data, we tested a model of the detection rates from \cite{georgieva2021} as well as the amended version presented here to recover the intrinsic detection efficiency $\eta$. We compared the results to values of $\eta$ retrieved from \textit{ex post}-validated, time-stamped data, where only trigger events were considered, that were separated from a previous event by more than the detector's holdoff time. We found that the original model may lead to a significant overestimate of $\eta$ compared to the \textit{ex post}-validated data. This overestimate increased systematically with a longer holdoff time and with increasing mean photon number per pulse. This systematic variation was strongly suppressed for the values of $\eta$ as recovered by the amended model and not observed using \textit{ex post}-validated data (`ground truth').\\Retrieving $\eta$ from experimental data, and corroborated by the simulations, we observed a significant decrease of $\eta$ with increasing mean photon number and repetition rate. For the devices considered here, this can largely be regarded as a function of the mean time of a signal detection to the previous detection event. We found that, for time intervals corresponding to detection rates of around 10~kHz or more, the quantum efficiency dropped significantly by more than 5~\% relative to its low-rate value, and by up to around 20~\% close to the maximum rate. We therefore conclude that the intrinsic detection efficiency of single-photon detectors in general cannot be regarded as having a constant value. This rather unexpected result is of utmost importance for quantum technologies, since consequently,  any calibration of $\eta$, in particular in view of high-rate applications, needs to cover a wide range of conditions and regimes to make sure that it does not miss unexpected behaviour that might affect the intended quantum technological application. For example, an unknown change of a detector's quantum efficiency under certain conditions might open a security loophole in quantum cryptography leaving the quantum key distribution (QKD) system vulnerable to attacks \cite{Diamanti2016,ETSIWP27}. For systems that employ multiple detectors, also a test and characterization of the difference in detection efficiency is necessary to avoid detection efficiency mismatch attacks \cite{Lo2014, Makarov2006}. Guidelines published by bodies such as the European Telecommunications Standards Institute (ETSI) \cite{etsi} include definitions and methods to measure the detection efficiency of SPAD-based single-photon detectors \cite{ETSI11}. Based on the results presented here, we strongly recommend an amendment to the guidelines to include the investigation of a possible variation of the quantum efficiency for any future calibration of devices used in quantum communications.

\section{Acknowledgments}
The work reported here was funded by the projects EMPIR 19NRM06 METISQ and EMPIR 20FUN05 SEQUME. These projects received funding from the EMPIR programme co-financed by the Participating States and from the European Union's Horizon 2020 research and innovation programme.\\The name of any manufacturer or supplier by PTB or INRIM in this scientific journal shall not be taken to be PTB’s or INRIM’s endorsement of specific samples of products of the said manufacturer; or recommendation of the said supplier.


\begin{thebibliography}{99}
	\bibitem{Sas11} M. Sasaki, M. Fujiwara, H. Ishizuka, W. Klaus, K. Wakui et al., \textit{Field test of quantum key distribution in the Tokyo QKD Network}, Opt Express \textbf{19}, 10387 (2011).
	\bibitem{Mig04} A. Migdall, \textit{Introduction to journal of modern optics special issue on single-photon: detectors, applications, and measurement methods}. J. Mod. Opt. \textbf{51}, 1265 (2004).
	\bibitem{Pee09} M. Peev, C. Pacher, R. Alléaume, C. Barreiro, J. Bouda et al, \textit{The SECOQC quantum key distribution network in Vienna}, New J Phys \textbf{11}, 075001 (2009).
     \bibitem{Pirandola2020} S. Pirandola, U. L. Andersen, L. Banchi, M. Berta, D. Bunandar, R. Colbeck, D. Englund, T. Gehring, C. Lupo, C. Ottaviani, J. L. Pereira, M. Razavi, J. Shamsul Shaari, M. Tomamichel, V. C. Usenko, G. Vallone, P. Villoresi, and P. Wallden, \textit{Advances in quantum cryptography}, Adv. Optics and Photon. \textbf{12} 1012 (2020). 
      \bibitem{Xu2020} F. Xu, X. Ma, Q. Zhang, H.-K. Lo, and J.-W. Pan, \textit{Secure quantum key distribution with realistic devices}, Rev. Mod. Phys. \textbf{92}, 025002 (2020).
	\bibitem{Med17} A. Meda, I. P. Degiovanni, A. Tosi, Z. Yuan, G. Brida and M. Genovese, \textit{Quantifying backflash radiation to prevent zero-error attacks in quantum key distribution}, Light: Science \& Applications e16261 (2017)
	\bibitem{Bru19} C. Bruschini, H. Homulle, I. M. Antolovic et al., \textit{Single-photon avalanche diode imagers in biophotonics: review and outlook}. Light Sci Appl \textbf{8}, 87 (2019).
	\bibitem{Kar17} M. A. Karami, and M. Ansarian, \textit{Neural Imaging Using Single-Photon Avalanche Diodes}, Basic Clin. Neurosci., \textbf{8}, 19 (2017).
	\bibitem{McC13} A. McCarthy, N. J. Krichel, N. R. Gemmell, X. Ren, M. G. Tanner, S. N. Dorenbos, V. Zwiller, R. H. Hadfield, and G. S. Buller, \textit{Kilometre-range, high resolution depth imaging using 1560 nm wavelength single-photon detection}, Opt. Express \textbf{21}, 8904 (2013).
	\bibitem{Ren11} M. Ren. et al., \textit{Laser ranging at 1550 nm with 1-GHz sine-wave gated InGaAs/InP APD single-photon detector}, Opt. Express \textbf{19}, 13497 (2011).
	\bibitem{Chan19} S. Chan, A. Halimi, F. Zhu et al., \textit{Long-range depth imaging using a single-photon detector array and non-local data fusion}. Sci Rep \textbf{9}, 8075 (2019). https://doi.org/10.1038/s41598-019-44316-x.
	\bibitem{Hadfield2009} R. H. Hadfield,  \textit{Single-photon detectors for optical quantum information applications}, Nature Photonics \textbf{3}, 696 (2009).
	\bibitem{Zha09} Q. Zhang, H. Takesue, T. Honjo, K. Wen, T. Hirohata, M. Suyama, Y. Takiguchi, H. Kamada, Y. Tokura, O. Tadanaga, Y. Nishida, M. Asobe, and Y. Yamamoto, \textit{Megabits secure key rate quantum key distribution}, New J. Phys. \textbf{11}, 045010 (2009). 
	\bibitem{Dix08} A. R. Dixon, Z. L. Yuan, J. F. Dynes, A. W. Sharpe, and A. J. Shields, \textit{Gigahertz decoy quantum key distribution with 1 Mbit/s secure key rate}, Opt. Express \textbf{16}, 18790 (2008).  
	\bibitem{Liao2017} S.-K. Liao, W.-Q. Cai, W.-Y. Liu, . Zhang, Y. Li, J.i-G. Ren, J. Yin, Q.i Shen, Y. Cao, Z.-P. Li, F.-Z. Li, X.-W. Chen, L.-H. Sun, J.-J. Jia, J.-C. Wu, X.-J. Jiang, J.-F. Wang, Y.-M. Huang, Q. Wang, Y.-L. Zhou, L. Deng, T. Xi, L. Ma, T. Hu, Q. Zhang, Y.-A. Chen, N.-L. Liu, X.-B. Wang, Z.-C. Zhu, C.-Y. Lu, R. Shu, C.-Z. Peng, J.-Y. Wang, and J.-W. Pan, \textit{Satellite-to-ground quantum key distribution}, Nature \textbf{549}, 43 (2017). 
	\bibitem{Val16} G. Vallone, D. Dequal, M. Tomasin, F. Vedovato, M. Schiavon, V. Luceri, G. Bianco, and P. Villoresi, \textit{Interference at the single photon level along satellite-ground channels}, Phys. Rev. Lett. \textbf{116}, 253601 (2016).
	\bibitem{Du18} Du, B., Pang, C., Wu, D. et al., \textit{High-speed photon-counting laser ranging for broad range of distances}, Sci. Rep. \textbf{8}, 4198 (2018).
	\bibitem{Had09} R. Hadfield, \textit{Single-photon detectors for optical quantum information applications}, Nature Photon. \textbf{3}, 696 (2009).
	\bibitem{Tit19} W. Tittel, \textit{Quantum key distribution breaking limits}, Nature Photon. \textbf{13}, 310 (2019).
	\bibitem{lopez2015} M. L\'opez, H. Hofer, and S. K\"uck, \textit{Detection efficiency calibration of single-photon silicon avalanche photodiodes traceable using double attenuator technique}, J. Mod. Opt. \textbf{62}, 1732 (2015)
     \bibitem{IEC61315} IEC 61315, \textit{Calibration of Fibre-Optics Power Meters} (International Electrotechnical Commission, Geneva, Switzerland), 3rd Ed. 
      \bibitem{goebel1997} R. Goebel, R. Pello, K.D. Stock, and H. Hofer, \textit{Direct comparison of cryogenic radiometers from the BIPM and the PTB}, Metrologia \textbf{34}, 257 (1997). 
	\bibitem{georgieva2021} 
		H. Georgieva et al., \textit{Detection of ultra-weak laser pulses by free-running single-photon detectors: Modeling dead time and dark count effects}, Appl. Phys. Lett. \textbf{118}, 174002 (2021)
     \bibitem{Brida2000} G. Brida, S. Castelletto, I. P. Degiovanni, M. Genovese, C. Novero, and M. L. Rastello, \textit{Towards an uncertainty budget in quantum-efficiency measurements with parametric fluorescence}, Metrologia \textbf{37}, 629 (2000).
     \bibitem{Brida2005} G. Brida, M. Genovese, M. Gramegna, M. L. Rastello, M. Chekhova, and L. Krivitsky, \textit{Single-photon detector calibration by means of conditional polarization rotation}, J. Opt. Soc. Am. B \textbf{22}, 488 (2005).
    \bibitem{Castelletto2006} S. Castelletto, I. P. Degiovanni, V. Schettini, and A Migdall, \textit{Optimizing single-photon-source heralding efficiency and detection efficiency metrology at 1550 nm using periodically poled lithium niobate}, Metrologia \textbf{43}, S56 (2006).
	\bibitem{Min15} M. G. Mingolla, F. Piacentini. A. Avella, M. Gramegna, L. Lolli, A. Meda, I. Ruo Berchera,  E. Taralli, P. Traina, M. Rajteri, G. Brida, I. P. Degiovanni, M. Genovese, \textit{Quantum and Classical Characterization of Single/Few Photon Detectors}, Quantum Matter \textbf{4}, 200 (2015).	
	\bibitem{lopez2020} M. L\'opez et al., \textit{A study to develop a robust method for measuring the detection efficiency of free-running InGaAs/InP single-photondetectors}, EPJ Quantum Technology \textbf{7}, 14 (2020).
	\bibitem{Diamanti2016} E. Diamanti, H.-K. Lo, B. Qi, and Z. Yuan,  \textit{Practical challenges in quantum key distribution}, npj Quantum Inf \textbf{2}, 16025 (2016). 
	\bibitem{ETSIWP27} [ETSIWP27] ETSI White Paper No. 27: \textit{Implementation Security of Quantum Cryptography. 
Introduction, challenges, solutions.} First edition – July 2018.
	   \bibitem{Lo2014}  H.-K. Lo, M. Curty, and K. Tamaki, \textit{Secure quantum key distribution}, Nature Photon. \textbf{8}, 595 (2014). 
	   \bibitem{Makarov2006} V. Makarov, A. Anisimov and J. Skaar, \textit{Effects of detector efficiency mismatch on security of 
quantum cryptosystems}, Phys. Rev. A \textbf{74} 022313 (2006).
	\bibitem{etsi} https://www.etsi.org/committee/qkd
\bibitem{ETSI11} ETSI Specification Document: GS/QKD-011, \textit{Quantum Key Distribution (QKD); Component characterisation: characterising optical components for QKD systems}, V1.1.1 (2016-05)
	
\end{thebibliography}

\end{document}